# Fluorescence Profile of Calabrian *Opuntia ficus-indica* Cladodes Extract: a Promising Low-cost Material for Technological Applications

Antonio Ferraro[a], Sephora Kamwe Sighano[b], Roberto Caputo[a,b], Franco Cofone[a], Giovanni Desiderio[a], Oriella Gennari[*a].

[a] Consiglio Nazionale delle Ricerche - Istituto di Nanotecnologia (CNR-Nanotec), 87036 Rende (CS), Italy
* E-mail: oriella.gennari@cnr.it

[b] Department of Physics, University of Calabria, 87036 Rende (CS), Italy

The autofluorescence of calabrian *Opuntia ficus-indica* bioactive extract upon UV illumination is explored by fluorescence spectroscopy enabling to investigate the typology and distribution of responsible molecules within green cladodes. The spectroscopic analysis of the extract shows a significative red emission, suggesting an abundance of chlorophylls in the sample. Such molecules show pronounced fluorescence in the visible range (400-800 nm) with a very large Stokes shift, when excited with UV light source. The fluorescence profiling is performed also in the case of polymers, such as poly(methyl methacrylate) (PMMA), poly(vinylpyrrolidone) (PVP) and polyvinyl alcohol (PVA), enriched with *OFI* extract with a signal improvement up to 40 times greater than the extract alone. This by-product fluorescent molecule can replace commercial dyes into several applications spanning from nano-optics to anti-counterfeiting and bioimaging.

## Introduction

The exploitation of agricultural crops and agri-food by-products represents a promising strategy to produce advanced materials based on renewable resources and implement circular model[1]. In particular, from plant pruning waste or parts of the plant with no intrinsic value it is possible to separate a variety of molecules such as pigments showing optical characteristics useful in other fields such as sensing, biochemical or imaging studies[2]. With these perspectives, plant fluorophores constitute a novel starting material with peculiar spectral properties complementing and overcoming some of the limitations in fluorescent proteins and dyes commonly used in bio-medical and optical application[3–5].

For their ability to prosper under stressful environments, cacti are the most conspicuous and characteristic plants of arid and semi-arid regions where are widely used to prevent soil erosion and desertification[6,7]. *Opuntia Ficus-Indica (L.) Mill.* (*OFI*), commonly known as prickly pear or cactus pear, is gaining interest across the world because it can grow where no other crops are able to do. *OFI* is the Cactaceae plant with the greatest economic and social, agronomic and ecological benefits in the world[6]. This is mostly due to the profitable production of its delicious fruits rather than other cactus parts, such cladodes (or simply nopal) which have been generally undervalued, being considered as pruning waste to exploit at most for feeding livestock[8–11]. The plant is native to Mexico and has subsequently spread in many world areas, such as South America, South Africa and, not least, the Mediterranean basin[12]. In Calabria, cultivation plantation of OFI covers an area of more than 50 hectares but it is widespread in all the territory as wild plants. OFI plant is characterized by three components which are flowers, prickly pear fruits and leaves (botanically called cladodes)[9]. Various parts of this plant are scientifically proven to have therapeutic potentials and are safe for human use[13–15]. Characterization of cactus pear species and cultivars has been widely performed by using morphological descriptors. Peel and pulp color vary from white to green, yellow, orange, red, pink, and purple. These colors are ascribed to the presence of various pigments that include chlorophyll, carotenoids, and betalains. The young cladodes of OFI, also known as nopalitos, contain functional polyphenols, like Ruthin, Iso-quercitrin, Narcissin and Nicotiflorin[6,16,17] and a series of polysaccharides with high molecular weight, such as mucilage and pectins having important functional properties, as rheological, medicinal and nutritional[15,18]. Indeed, the beneficial advantages of *OFI* extract was well recognized in folk medicine for the treatment of wound healing, obesity and edema[18,19]. Beyond the notable functional and nutraceutical values, development and application of biomarkers based on vegetable extracts containing natural pigments is becoming of great interest for their use in bio-medical and optical fields[4,5]. To this aim, it is essential to increase the knowledge of OFI cladodes for optimizing their exploitation as an innovative ecofriendly material. Furthermore, fluorescent materials are receiving ever-increasing attention and the range of application areas in which they have been applied is seemingly ever-expanding. In addition to being used as dyes, fluorescent materials are attractive for their potential application in many fields, such as in optical devices and brighteners, photo-oxidants, coatings, chemical and biochemical analyses, solar traps, anti-counterfeiting labels, drug tracers, information storage[20], sensing and imaging[3].

Considering our research in the design of advanced materials and the valorization of agri-food waste products, the aim of this study is to establish an extraction strategy to obtain a fluorescent offprint with peculiar optical features from *Opuntia ficus-indica* cladodes. OFI extract, composed exclusively by chlorophyll, was successfully added to polymethylmethacrylate (PMMA), Polyvinylpyrrolidone (PVP) and polyvinyl alcohol (PVA). Fluorescence spectroscopy was used to study and characterize the strong red emission upon UV excitation. Stability of polymeric solutions were also tested over time to prove the resistance of fluorescent materials under environmental stressors. The results obtained for *OFI* extract in toluene were attractive and reliable over a long period. While chlorophyll fluorescence has been extensively utilized as a non-invasive tool in the assessment of photosynthesis[21], plant physiology[22], $CO_2$ assimilation and linear electron flux[2] techniques, it is noteworthy that fluorescent extracts obtained from waste products have rarely been utilized for enriching polymeric materials and conferring fluorescence properties upon them. This presents an opportunity to explore the potential of low-cost products with intriguing optical properties, offering promise for various technological applications[23,24].

## Results and discussion

The agri-food materials are complex matrices comprising a variety of molecules, metabolites as well as several other co-extracted compounds that could interfere in qualitative analysis[25,26]. In the case of cladodes, water, carbohydrates, and proteins are the main chemical components. The carbohydrates, with structural and storing functions, were removed by cutting out the internal part of the cladodes, *medulla,* allowing also to eliminate fibers contribution; water was eliminated through lyophilization. The choice of more mature leaves guarantee to reduce protein content and deriving interferences, higher in young cladodes related to the improved metabolic activity[8]. The powder obtained after lyophilization process, Figure 1c, was observed using a scanning electron microscope (SEM) which shows what happened to the treated sample: the effectiveness of the extraction process is confirmed by the fibrous residue which appears completely colorless. The inner tissue, called chlorenchyma, consists of green plastids observable before the procedure and exhaustively removed at the end of the extraction, as shown in Figure 1d and -e, respectively.

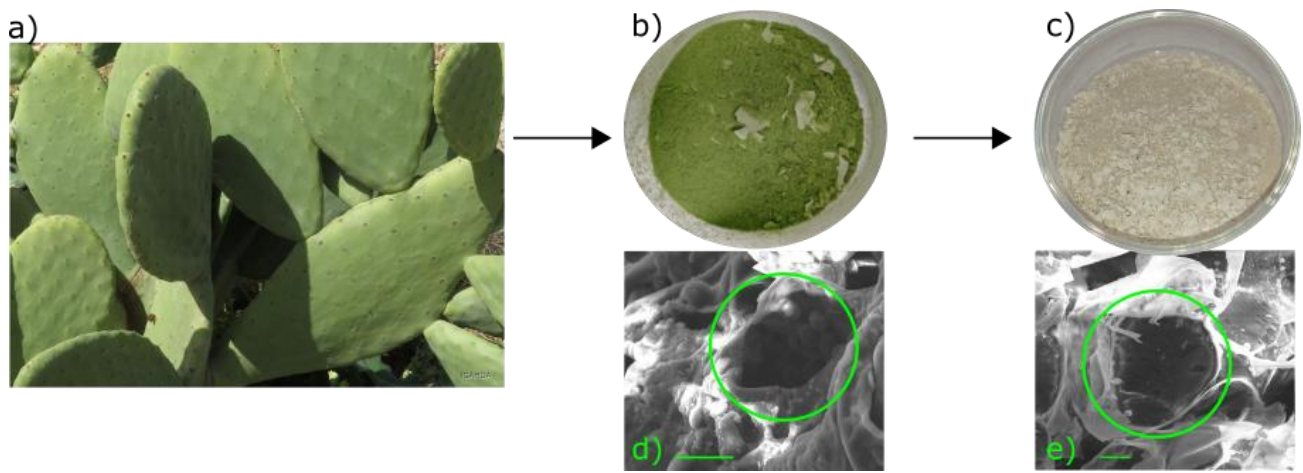

*Figure 1: (a) Photography of OFI cladodes; powder before (b) and after (c) extraction procedure. Related SEM images (different magnification) of two distinct phases of chloroplast state: (d) before with the presence of plastids (inner circle zone) and (e) after extraction process. Scale bar: 20 µm.*

While the possibility of extracting and characterizing polyphenolic structures is widely discussed[4,6,11,13,15–19], no satisfactory literature currently exists regarding the extraction and separation of other types of fluorescing substances from the *Opuntia ficus-indica*.

Here, we propose a simple extraction approach letting to obtain a green residue made up exclusively by fluorescent pigments, named chlorophyll. Fluorescence from intact plant tissues is the result of multi-component emission of light from several substances present in cells, simplifying the starting sample is thus essential to obtain cleaner and reproducible fluorescence signals.

In case of our matrix, the exclusion of the inner part of cladodes – *medulla* – and the complete removal of water through lyophilization allowed us to work on a simplified sample.

The following steps let us to obtain a green extract composed exclusively by chlorophyll molecules[27].

The numerous chloroplasts dispersed in the cytosol around the vacuoles (see Figure 1d) in *OFI* cladode powder justify the red fluorescence[28], that is only emitted by Chlorophylls[29–31]. All chlorophyll molecules possess extensive conjugated bonds in the form of a porphyrin ring that are responsible of fluorescence. It is observed that the fluorescence spectra are characterized by a leading Gaussian peak in the red band followed by a shoulder into the NIR spectral band[32].

Fluorescence analysis performed on the starting power dissolved in water, see Figure 2a, demonstrates the need of extraction purification to enhance the intensity of fluorescence signal[3,33]. Indeed, the spectra of purified OFI extract in toluene, reported in Figure 2b, was recorded decreasing the integration time to 1 sec and shows well defined fluorescence peak centered at 680-685 nm.

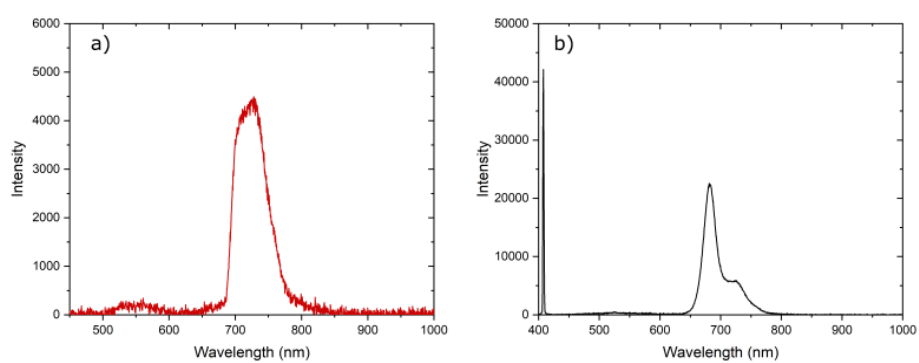

*Figure 2: Fluorescence measurement performed using an excitation wavelength ($\lambda_{ex}$) of 405 nm with power of 35 mW on OFI powder obtained from lyophilized parenchyma before extraction procedure (a) in water and (b) in toluene. Integration time of 5 sec and 1 sec respectively. Different y-axis scale.*

To ensure the complete breaking of plastids[34] (see Figure 1d and 1e) in which fluorescence emitting substances are contained, a preliminary step using methanol for the powder rehydration was applied and necessary to perform the following extraction. Methanol guaranteed the function of solubilizing the more polar pigments and separating them from the *OFI* matrix[29,35,36].

The fluorescence spectrum (see Figure 2b) shows the absence of the interferences from other fluorescing substances, clearly indicating that the proposed procedure was useful and efficient; the acetone/ethanol/diethyl ether represented a good choice allowed to obtain a clean extract, avoid the interferences of phenolic fraction removed by methanol phase and obtain chlorophyll selectively. The analytical determination of plant fluorophores by fluorescence has two big advantages, namely great sensitivity and selectivity, due to the unique property of autofluorescent molecules excited by a specific wavelength ($\lambda_{ex}$) and emitting radiation of a specific wavelength ($\lambda_{em}$)[33].

The reliability and the multipurpose capability of the *OFI* extract was demonstrated by mixing it with different polymers, such as PMMA, PVP and PVA to a final concentration of 15% w/w. See **Experimental** section for further information. The solutions were excited at a wavelength of 405 nm and the resulting fluorescent emission was recorded.

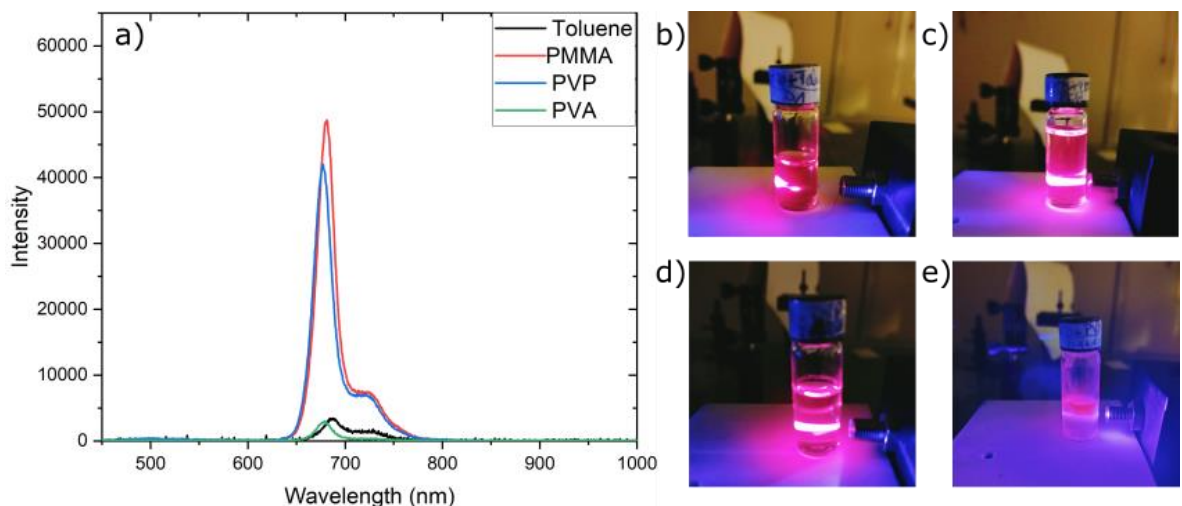

*Figure 3: a) Fluorescence spectra of the Opuntia ficus-indica (OFI) extract dissolved in toluene (black line) and of polymers (PMMA, PVP and PVA) enriched with OFI extract. Excitation wavelength of 405nm and power of 5 mW; integration time of 1 sec. Photographs of emitted fluorescence of the analyzed samples (b) OFI extract in toluene; (c) OFI extract in PMMA; (d) OFI extract in PVP and (e) OFI extract in PVA.*

As in case of the *OFI* extract alone and mixed with polymers, upon excitation at 405 nm, fluorescence results reveal pronounced features in the visible range (400-800 nm) with a very large Stokes shift of 280 nm that is usually difficult to obtain with chemically synthesized dyes. As shown in Figure 3, the emitted fluorescence was strongly observed in the red region with a distinct bimodal emission with maxima at 680-685 nm, corresponding to the monomer form, and 720-730 nm, depending of Chlorophyll aggregates[37–39]. By comparing the fluorescence spectrum of *OFI* extract with those of polymer enriched with it, the intensity obtained in the case of PVA solution (green line) was comparable with that recorded for *OFI* toluene extract as it was (black line). In case of PMMA (red line) and PVP (blue line) enriched with extract, the signal intensity was clearly enhanced up to a fourfold increase (see Figure 3). The emission band in these polymers was considerably increased due to the existence of a mixture of components with synergistic effect in enhancing the fluorescence properties. The emission of many fluorophores is sensitive to the polarity of their surrounding environment[40]. As the medium polarity is decreased, the emission spectrum gets reduced. Conversely, the increase of solvent polarity generally results in enhancement of the emission spectrum. According to this, the results show clearly that the highest fluorescence peak is provided by PMMA and PVP, while a poor effect is observed in case of PVA. The polymers PMMA, PVP, and PVA are polar dielectrics thanks to the ester (PMMA), amide (PVP) and hydroxyl (PVA) groups[41], where the ester exhibits stronger polarity. The carbonyl group, common in the chemical structure to both PMMA and PVP, works as an auxochrome determining a hyperchromic effect of the emitted signal intensity[42]. The presence of the ester group, for a conjugation effect, significantly increases the intensity in respect to the amide group which explains the difference between PMMA and PVP[41]. The differences in the interacting molecular species provides a different aggregate state of chlorophyll thus determining a hyper- or hypochromic effect, depending on the polymeric macromolecule. The locations of the fluorescence maxima were the same, but the peak heights were significantly different. The existence of chlorophyll aggregates, detected by additional fluorescence as a shoulder at ≈720 nm, suggests a small contribution of aggregated forms. These results then indicate that the aggregation of chlorophyll in either macromolecule is strongly affected by the nature of macromolecules[37,43,44]. Figure 3b-d qualitatively confirms the fluorescence intensity behavior by showing the emission color of OFI extract in toluene, in PMMA, in PVP and in PVA respectively under UV excitation.

**Stability of fluorescent materials**

The stability of chlorophyll poses a significant challenge due to the ease how the molecule readily degrades into several metabolites under normal conditions[45].

The stability of the toluene *OFI* extract and of the polymeric materials enriched with it was tested. In particular, the fluorescence signal of samples after six months since their preparation was measured and compared with the first analysis, see Figure 4. For all solutions, a reduction in peak height, indicative of signal intensity, was observed while maintaining peak quality and profile. A reduction of 27% was obtained for the case of OFI extract alone, 43% for PMMA, 20% for PVP and 41% for PVA. The intensity reduction of the peak is attributable to a loss in title which normally occurs during time, and which sets the maximum stability of a solution at six months.

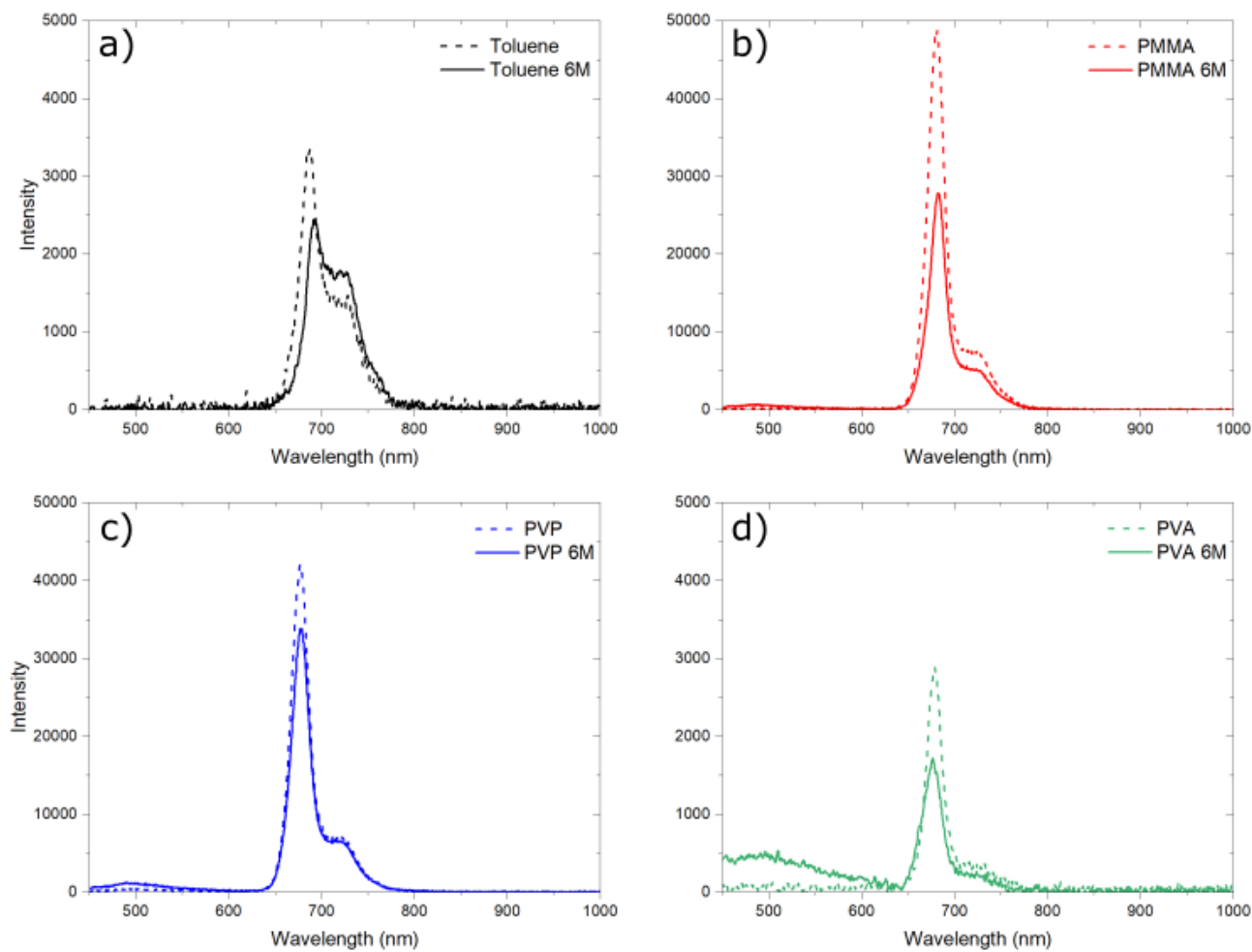

*Figure 4: Fluorescence spectra of the Opuntia ficus-indica (OFI) extract dissolved in toluene (a) and of PMMA (b), PVP (c) and PVA (d) enriched with it at time 0 (dashed line) and after 6 months (continuous line). Different Y axis scale between a, d and b, c.*

Beyond time, chlorophyll fluorescence response under the influence of different external factors was studied. To mimic environmental conditions, the *OFI* extract in toluene was saturated with carbon dioxide, $CO_2$, at a flow rate of 0.15 mL/min and fluorescence spectrum was acquired after 1h. No relevant difference was observed in intensity and profile of the fluorescent peak. Probably, it was due to the inability of $CO_2$ to modify the pH value of the toluene sample. In fact, $CO_2$ will increase aqueous solution acidity because it reacts with water to form carbonic acid, $H_2CO_3$, a weak acid but this is limited to aqueous environment.

Therefore, to prove the stability of the extract in a protic solvent subjected to $CO_2$ saturation, the residue of the extraction was recovered with water and the pH was measured. A pH meter (Multi Meter MM41, Crison Instruments, Spain), calibrated with pH 4.01, 7.00 and 9.21 buffer solutions, was used. The initial pH value measured for the OFI extract in water was 4.30. The production of carbonic acid during the experiment did not alter drastically the pH solution that was 4.08 after 1h of $CO_2$ insufflation. After, the difference observed in the color did not seem consistent (see Figure 5b), and a fluorescence analysis showed the typical profile of chlorophyll (Figure 5d – red line). After 8 weeks, the extract showed a slightly loss of color, see Figure 5c. It is important to underline that in the case of *OFI* extract in water, a completely natural degradation was observed due to the passage of time, given that the sample was stored at a controlled temperature of 4 °C and protected from light.

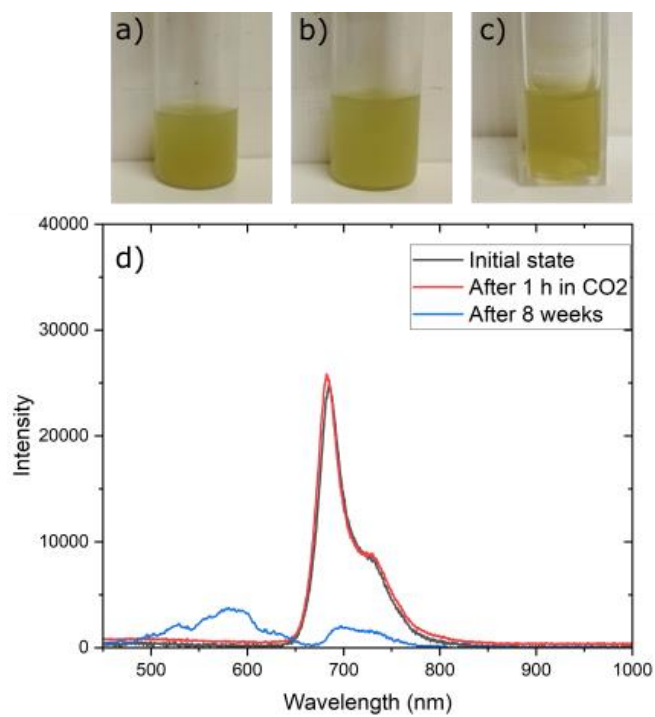

*Figure 5: Photographs of OFI extract in water a) initial solution, b) after 1h of $CO_2$ insufflation, c) after 8 weeks. d) fluorescence spectra of initial solution (black line), after 1 hour of insufflation with Co2 (red line) and after 8 weeks (blue line). Laser power 35mW and integration time of 1 sec for black and red line while 5 sec for blue line due to low signal.*

The difference of fluorescence emission is reported in Figure 5d. As mentioned above, standard chlorophyll fluorescence emission occurs in the wavelength range of 650 and 800 nanometres with two typical peaks corresponding to red (680-695 nm) and far-red (720-730 nm) wavelengths. Red fluorescence (F685) occurs in photosystem (PS) II, and far-red (F720) in PS I[46]. The ratio of red to far-red fluorescence (F685/F720) is used to estimate environmental stress[22,47,48] and are correlated with different state of chlorophyll aggregation[37,44,49,50]. As can be seen in Figure 5d (red line), the fluorescence spectra is not affected meaning that the OFI extract is stable also in water, probably because the pH variation was minimal[37,49,50]. The extract was stored for 8 weeks, and then the spectra showed new features of chlorophyll peak. It is impossible to resolve the two peaks (the main one and the shoulder) in the spectral range 650-800 nm, thus explaining the difficulty to estimate the F687/F760 ratio. At 680 nm, an 84.5% reduction of intensity is evident, and the fluorescence peak appears flattened and degraded. The spectrum points out the presence of new fluorescents in the emission range from 500 to 650 nm, with spectral characteristics compatible with degradants, such as pheophytins. The conversion of chlorophyll to pheophytin justifies the colour change from bright green to olive-yellow (see Figure 6).

Strong pH influences on the chlorophyll fluorescence were observed in literature, being capable of modulating the aggregation state of chlorophyll monomers[49]. Therefore, we studied the spectral behaviour of chlorophyll in different living environments to understand how the pH change enabled the variation in emission spectra and influenced the state of chlorophyll aggregation. Firstly, the pH range of chlorophyll solution was controlled from 4.30 to 11.60 by adding different amounts of $NH_4OH$ and the related fluorescence spectra are shown in Figure 6a, b, respectively. In the latter case, at pH 11.60, the two peaks matched with a visible reduction of the intensity at 680 nm. In this condition, chlorophyll monomer state was equal to the amount of chlorophyll aggregates. By adding $CH_3COOH$, the pH was restored to a pH value around 4.30 and fluorescence spectrum, see Figure 6c, suggested a predominant contribution of aggregated chlorophyll forms in the medium. The sample was successively alkalinized with small amounts of $NH_4OH$ until pH around 11.60 and the corresponding fluorescence spectra, see Figure 6d, evidenced an improvement of the intensity peak at 680 nm, corresponding to the monomeric state, with a small contribution of aggregate forms at 720 nm. The pH change induced a gradual decrease of the fluorescent emission; a hypochromic effect at 680 and 720 nm has been clearly observed in Figure 6c, d, and resulted in chlorophyll decrease with a simultaneous appearance of new fluorescent derivatives.

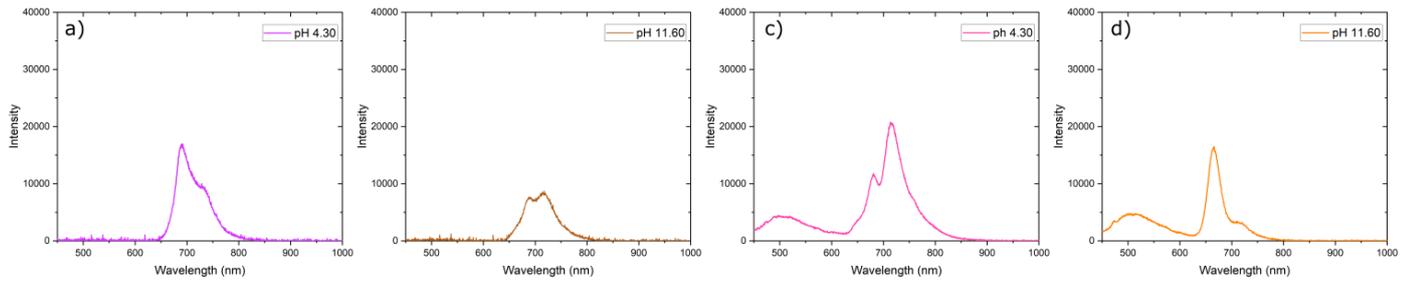

*Figure 6: pH influences on the chlorophyll fluorescence. a) initial pH= 4.30 increased to b) pH=11.60; decreased to c) pH=4.30 and increased again to d) pH=11.60.*

The fluorescence performance of samples reported in Figure 6c and d was monitored for 8 weeks and the results reported in

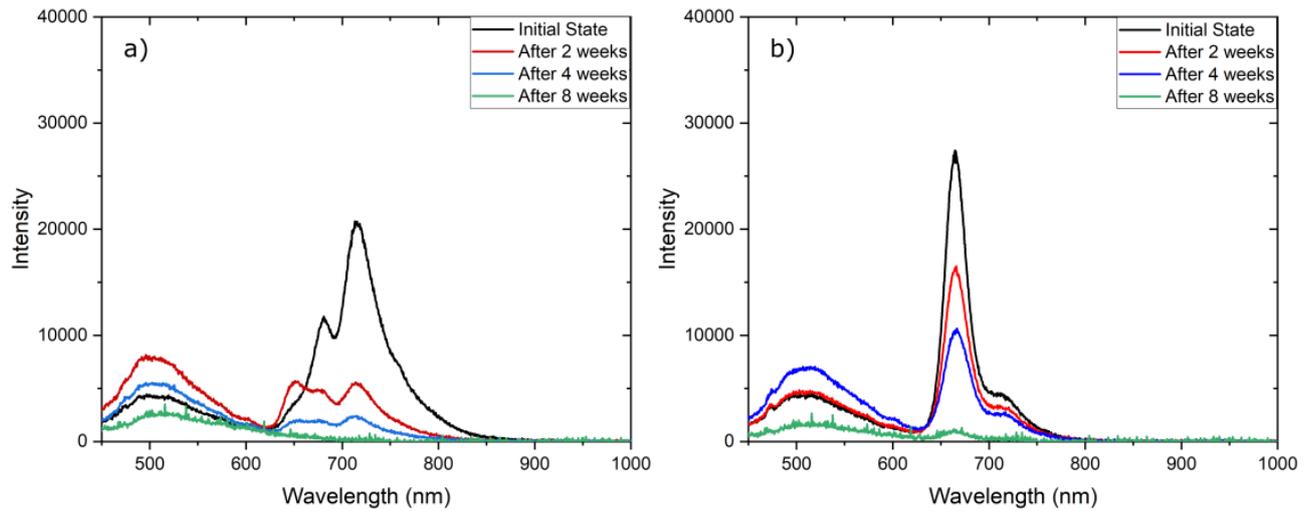

Figure 7a and b, respectively, evidenced the emission of other metabolites as result of structural damages caused by environmental stressors. As can be seen in

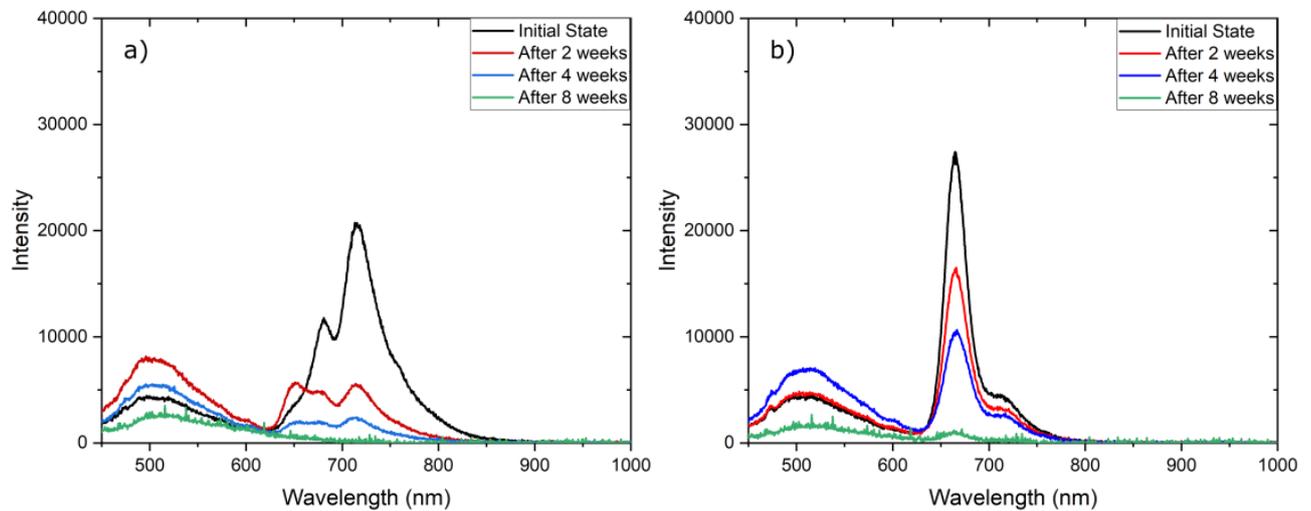

Figure 7a and b, a linear reduction of the chlorophyll peaks, at 680 and 720 nm, was observed with the simultaneous appearance of peaks in the emission range between 450-600 nm. The peaks mainly observed were related to lipofuscins (400-450 nm) and pheophytins (550-600 nm)[5]. The color of samples was completely lost and there is general agreement that the main cause of discolouration of green extracts is the conversion of chlorophylls to pheophytins, a natural initial, but irreversible, step in

chlorophyll degradation[37], promoted by the pH change (see

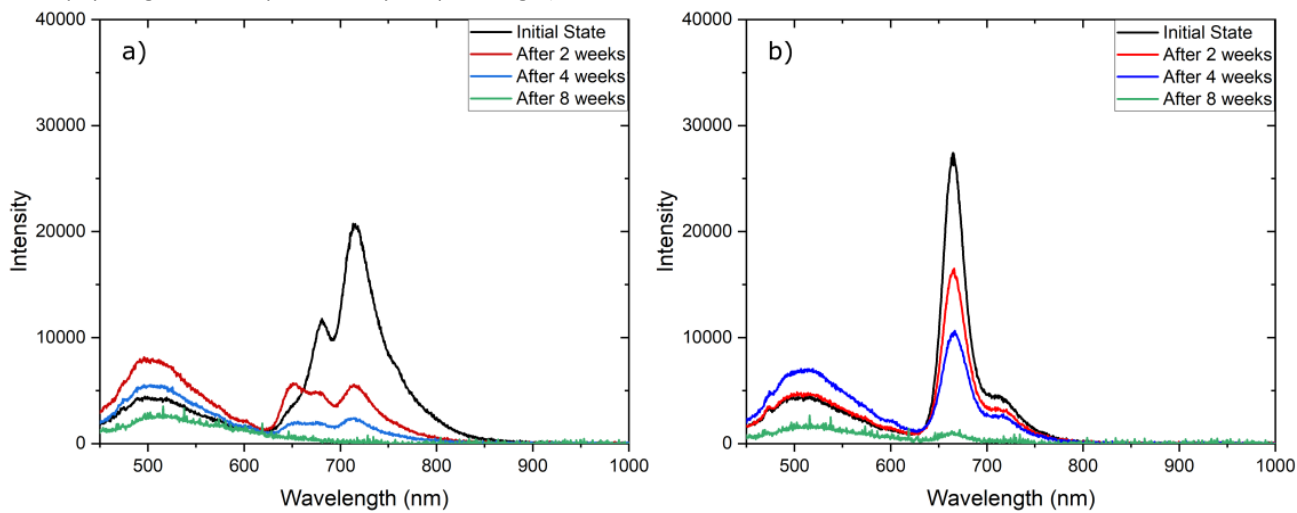

Figure 7).

The profiling of degradants in the emission range 450-600 nm is subject to further investigation by our research group to define the nature of the compounds generated by the chlorophyll metabolism.

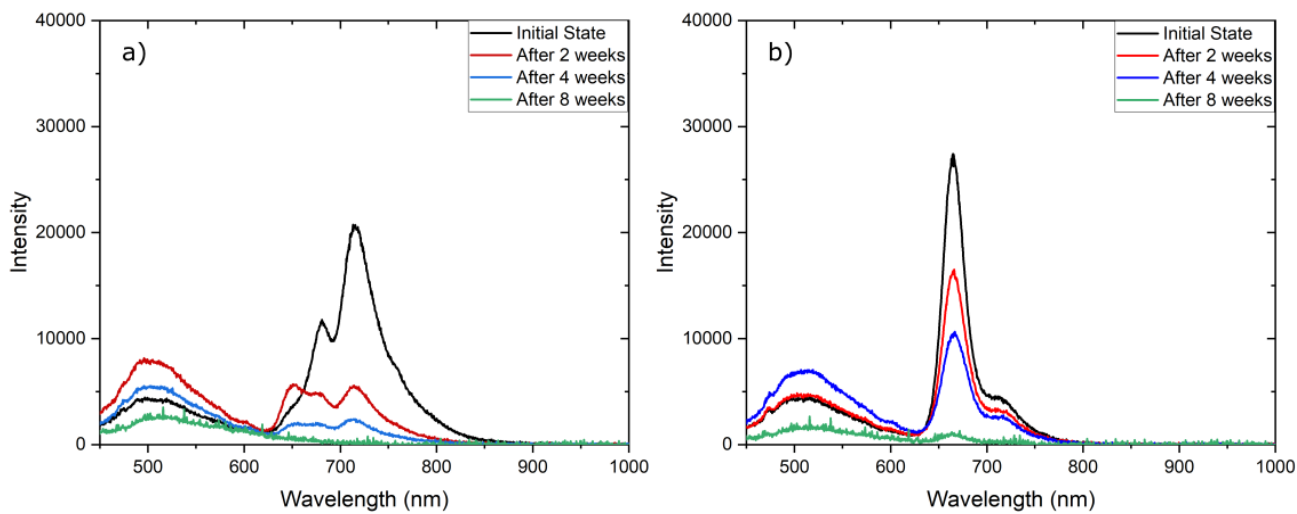

*Figure 7: pH-induced degradation of Chlorophyll in the aqueous medium: the changes in the fluorescence spectra followed by pH change for 8 weeks. Excitation Power of 35mW and integration time of 5 seconds. a) pH of 4.30 b) pH: 11.60*

The experiment was repeated at least three times and the same behavior of the extract was observed.
The same protocol was also performed in case of the *OFI* extract in toluene but no variations in fluorescence behavior were measured. This result confirmed the stability of the toluene extract and of the polymeric solutions over time, if compared with aqueous solutions.

## Experimental

**OFI samples and reagents**

Cladodes from autochthonous cultivars of *Opuntia ficus-indica* (L.) Mill., 2-3 years old, were recovered from wild plantation located in the northern tyrrenian area of Calabria, in the period from May to June, see Figure 1a. Polymer solutions were employed in order to demonstrate the wide applicability of the fluorescent extract: poly(methyl methacrylate) (PMMA) (MW ~350,000), poly(vinylpyrrolidone) (PVP) (MW ~340,000) and polyvinyl alcohol (PVA) (MW 85,000-124,000).
All chemicals and reagents used during extraction and analysis were analytical grade and purchased from Sigma-Aldrich (Milan, Italy).

**Extraction of chlorophylls from cladodes and preparation of *OFI* extract**

After collection, all samples were washed with deionized water and cleaned from thorns; the green outer part of cladodes (*parenchyma*) was peeled off and separated from the inner part (*medulla*), cut in small pieces and homogenized, see Figure 1b. A portion of 50 grams was taken and freeze-dried using a Freezone 2.5 model 76530 lyophilizer (Labconco Corp. Kansas City, MO, USA) for 48 hours to obtain a powder. Firstly, three grams of sample powder was extracted with Acetone/Ethanol 2:1 (v/v) mixture[34] and, finally, with diethyl ether until the residue was colorless, see Figure 1c. Each extraction was performed using an ultrasonic bath followed by filtration through a Whatman No.1 filter paper. The phases were combined and evaporated under vacuum at 40 °C. The procedure was carried out in triplicate on lyophilized powder. The dried residue was dissolved in toluene and analyzed by fluorescence spectroscopy.

**Fluorescence Spectroscopy**

Fluorescence measurements were performed for both the *OFI* extract alone and the polymeric solutions enriched with it. With this aim, polymeric solutions of PMMA, PVP and PVA were prepared by dissolving 1g of polymer in toluene, ethanol, and water respectively to a final concentration of 10% w/w. An aliquot of *OFI* extract in toluene was mixed with the polymer solution to a final concentration of 15% (w/w) of extract in polymer. For all fluorescence measurements the laser excitation wavelength was fixed at 405 nm with a beam diameter of 3 mm (Micron Laserage model LDM405.120.CWA). For all measurement the excitation power was 5 mW apart the one related to OFI pieces and OFI in water (Figure 2, Figure 5) that was 35 mW, due to low fluorescence signal. Integration times (indicated in the Figures) of 1 sec and 5 sec are used depending on the sample. All the spectra were recorded using an optical spectrometer (FLAME, Ocean Optics).

**Characterization by Scanning Electron Microscopy**

The morphology properties of the OFI powder were analyzed using a Scanning Electron Microscopy (Quanta FEG 400F, Fei Company, USA). Before the analysis, the samples were fixed on an aluminum specimen holder with carbon tape and then the mounted samples were sputter coated with graphite. Views of the OFI section samples were taken to obtain the micrographs.
The conditions of the analysis were high vacuum, 15 KV electron acceleration voltage, and secondary electron mode.

# Conclusions

The recovery of chlorophylls from food/plant by-products is an intriguing approach for improving the sustainability of food production and of enormous interest in food supply chain waste valorization. In this study, we set up an extraction protocol to obtain fluorescent chlorophyll from calabrian *Opuntia ficus-indica* cladodes, which are considered as a by-product of agroindustry. The extraction procedure was simple and guaranteed obtaining chlorophylls selectively. The fluorescence performance of the extract was explored by fluorescence spectroscopy and the characterization supported the great potential application of this natural fluorophore as additive in PMMA, PVP and PVA solutions. As in case of the *OFI* extract alone and mixed with polymers, upon excitation with UV source, fluorescence results reveal pronounced features in the visible range (400-800 nm) with a very large Stokes shift of 280 nm that is usually difficult to obtain with chemically synthesized dyes.
The stability of the fluorescence signal to external factors as $CO_2$, time, and pH variations, was comforting and promising for the possibility of using low-cost materials, as cladodes, to obtain molecules of technological potential in a very large field of applications. Owing to these interesting properties, fluorescent materials can be applied in many fields, including fluorescent sensors, probes, biological imaging, light-emitting diodes, and organic electronic devices.

# Author contributions

O.G. conceived the idea, designed the experiments, and prepared the manuscript.
A.F. and S.K.S. performed the fluorescence experiments.
A.F. analyzed the fluorescence data and contributed to the preparation of the paper.
F.C. contribute to extraction experiments.
G.D. performed SEM analysis.
R.C. contributed to the revision of the paper.
All authors have given approval to the definitive version of the manuscript.

## Conflict of Interest

There are no conflicts of interest to declare.

## Acknowledgements

O.G. acknowledges financial support from the project "Tech4You - Technologies for climate change adaptation and quality of life improvement" - ECS00000009, in the framework of PIANO NAZIONALE DI RIPRESA E RESILIENZA (PNRR).

A.F. acknowledges the financial support from project "PRIN 2022 2022T3B4HS_PE11 - Multi-step optical encoding in anti-counterfeiting photonic tags based on liquid crystals (PHOTAG)" financed in the framework of PIANO NAZIONALE DI RIPRESA E RESILIENZA (PNRR).

S.K.S. acknowledges the financial support from project "D.M. 10 agosto 2021 n. 1061 - PON 2014-2020 Dottorati di Ricerca su tematiche Green e dell'Innovazione" financed by "Ministero dell'Università e della Ricerca".

All authors acknowledge Dr. A. Bozzarello for his administrative support and management in the projects.